\documentclass[prb,nobibnotes,floatfix,superscriptaddress,twocolumn]{revtex4-2}
\usepackage{amsfonts,amsmath,graphicx,epsfig,latexsym}
\usepackage{subfigure}
\usepackage{graphicx}
\usepackage{color}
\usepackage{natbib}
\usepackage{multirow}
\usepackage{latexsym,amssymb}
\usepackage{amsmath}
\usepackage{soul}
\usepackage{ulem}
\usepackage{array}
\usepackage{siunitx}
\newcolumntype{d}[1]{D{.}{.}{#1}}
\usepackage[linktocpage,bookmarksopen,bookmarksnumbered]{hyperref}
\usepackage[dvipsnames]{xcolor}
\usepackage{placeins}
\usepackage{comment}
\usepackage{braket}

\usepackage{hyperref}

\usepackage{multirow}

\begin{document}

\title{Systematic dynamical mean-field theory study of 3$d$ perovskite oxides with uniform Coulomb interactions}

\author{Antik Sihi}
\affiliation
{Department of Physics and Astronomy, West Virginia University, Morgantown, WV, USA}

\author{Caden Ginter}
\affiliation
{Department of Physics and Astronomy, West Virginia University, Morgantown, WV, USA}

\author{Kristjan Haule}
\affiliation{Center for Materials Theory and Department of Physics and Astronomy, Rutgers University, Piscataway, NJ 08854, United States}

\author{Subhasish Mandal}
\email{Contact author: subhasish.mandal@mail.wvu.edu}
\affiliation
{Department of Physics and Astronomy, West Virginia University, Morgantown, WV, USA}

\begin{abstract}
Strongly correlated transition-metal perovskite oxides pose a fundamental challenge for first-principles electronic-structure theory and for large-scale, data-driven materials discovery. While density functional theory combined with dynamical mean-field theory (DFT+DMFT) provides a quantitatively accurate description of such systems, its high-throughput application is hindered by the need to determine material-specific Coulomb interaction parameters ($U$). First-principles approaches such as the constrained random phase approximation predict a highly nonlinear and non-transferable evolution of the interaction strength across chemically similar ABO$_3$ perovskites, reinforcing the conventional belief that transferable interaction parameters are incompatible with correlated-electron calculations. Here we show that this paradigm does not extend to the large-energy-window embedded DMFT (eDMFT) framework. In contrast to conventional downfolded DFT+DMFT approaches based on material-dependent Wannier orbitals, the present eDMFT implementation employs highly localized orbitals and treats electronic correlations and screening self-consistently within the same many-body framework. As a result, spectral properties are governed primarily by the dynamical self-energy rather than by static interaction-induced energy shifts, leading to a markedly reduced sensitivity to the precise value of the interaction parameter. Recent constrained-eDMFT calculations demonstrated that, for broad classes of $3d$ transition-metal oxides, the self-consistently screened Coulomb interactions naturally fall within relatively narrow ranges for correlated metals and insulators. Motivated by these findings, we implement a high-throughput eDMFT framework employing physically derived interaction values of $U=6$~eV for metals and $U=10$~eV for insulators together with {\it exact} double counting. We test this framework using systematic high-throughput eDMFT calculations for ABO$_3$ compounds (A = Ca, Sr, La; B = V--Ni) and benchmark the resulting densities of states and momentum-resolved spectral functions against photoemission experiments, where we find overall excellent agreement. Our results establish that charge self-consistent eDMFT enables robust, parameter-tuning-free high-throughput many-body calculations for correlated oxides, opening a practical pathway toward predictive electronic-structure databases for strongly correlated materials.

\end{abstract}


\maketitle

\section{Introduction}
Transition-metal-based perovskite oxides have long served as a central platform for condensed matter research due to their rich array of emergent phenomena, including unconventional superconductivity, magnetism, colossal magnetoresistance, multiferroicity, and photovoltaic responses \cite{perovskite_intro1,perovskite_intro2,perovskite_intro3,perovskite_intro4,perovskite_intro5,perovskite_intro6,Imada1998}. In particular, single perovskites with chemical formula ABO$_3$, where alkaline or rare-earth ions occupy the $A$ site and transition-metal (TM) ions occupy the $B$ site, provide a prototypical setting for studying the strong interplay among electronic, spin, and lattice degrees of freedom \cite{cox2010transition,dagotto2005complexity,Okamoto-Nature, Okamoto_PhysRevB.70.075101}. The partially filled, crystal-field-split TM $3d$-orbitals often give rise to strong on-site Coulomb interactions that dominate their low-energy electronic structure \cite{tokura2003correlated}. Consequently, many ABO$_3$ perovskites belong to the class of strongly correlated materials, exhibiting phenomena such as correlation-driven metal--insulator transitions, Hubbard band formation, spectral weight redistribution, and pronounced spin and orbital fluctuations \cite{Imada1998,georges1996dynamical, oles2012fingerprints, correlation_ligand_n9qh-6739}. These effects substantially modify both spectral and transport properties and place severe limitations on the conventional electronic-structure descriptions, as obtained in the density functional theory (DFT) \cite{RevModPhys.87.897,dft_fail1_cohen2012challenges}.

To address these challenges, several beyond–DFT approaches have been developed over the past three decades, such as meta-GGAs~\cite{mbj,mbj-jp,r2-scan,bansil-ph}, hybrid functionals~\cite{becke_new_1993,doi:10.1063/1.1564060,PhysRevB.83.235118,doi:10.1063/1.1564060,perdew_rationale_1996,SM-mtu1}, and the GW approximation~\cite{hedin_new_1965,refB4,hybertsen_electron_1986,Hybertsen_Na,KIM2019427,PyGW}. For perovskites, one of the most popular approaches is the DFT+$U$ method, which provides a computationally efficient extension of DFT and has been widely applied to correlated materials~\cite{Hub,Anisimov_1997,liechtenstein1995density,SIC_zaanen1988can}.
However, the static nature in DFT+$U$, enforces infinite quasiparticle lifetimes, can artificially open gaps through long-range order, and precludes a faithful description of temporal electronic correlations. As a result, essential features of correlated electron physics—such as quasiparticle renormalization, and temperature-dependent spectral evolution observed in photoemission—are inherently inaccessible within these frameworks~\cite{martin2016interacting}.  Dynamical mean-field theory (DMFT) overcomes these limitations by explicitly treating local electronic correlations in time, encoding correlation effects through a frequency-dependent self-energy~\cite{georges1996dynamical,kotliar2006electronic}. When combined with density functional theory, DFT+DMFT provides a nonperturbative framework in which dynamical correlations are captured on equal footing with the underlying band structure, enabling a quantitatively accurate description of the electronic structure of strongly correlated perovskite oxides and related systems \cite{PhysRevLett.94.026404,PhysRevLett.109.156402,PhysRevB.85.094505,mandal-PRL,PhysRevB.98.075155,turan-review,TMO1-SM,TMO2-SM,PhysRevB.98.075155, sihi_ce_singh2023evidence,sihi_prb_singh2020coexistence,sihi2020detailed,sihi2022exploring,sihi2023track,Antik-EPC}. Consequently, dynamical electronic correlations are now recognized as essential for a quantitative understanding of the electronic structure and transport properties of ABO$_3$ perovskites with partially filled 3{\it d}-electrons \cite{tokura2000orbital,haule2017mott,medarde1997structural,LaNiO3_ARPES_PhysRevB.92.245109, PhysRevB.90.125114, PhysRevB.102.045146, PhysRevB.98.075155,PhysRevB.107.045147, KH_lvo_PhysRevB.90.075136,dutta2018studying}.

Beyond individual materials studies, the accuracy of correlated electronic-structure methods has become increasingly important for emerging data-driven materials discovery. Recent advances in artificial intelligence and machine learning have accelerated efforts to explore vast materials spaces using databases of first-principles calculated properties \cite{jain2013materialsproject,curtarolo2012aflow,curtarolo2013aflowlib,JARVIS, Choudhary_ALIGNN2021, Mandal-alignn,butler2018machine,pati-ML}. While this paradigm has proven highly successful for weakly correlated materials—where DFT-based approaches are generally reliable—it remains fundamentally constrained for strongly correlated systems. Databases constructed primarily from DFT or DFT+$U$ calculations often yield qualitatively incorrect electronic structures for materials such as ABO$_3$ perovskites, severely limiting their applicability for data-driven discovery of correlated electron materials \cite{TMO1-SM,hafiz2018high, GW_NM_varrassi2025automated,uhrin2025machine}.

A central practical obstacle in large-scale DFT+DMFT calculations is the determination of the on-site Coulomb interaction parameter \(U\). Within static frameworks such as DFT+$U$, it is widely accepted that \(U\) must be strongly material dependent~\cite{PhysRevB.58.1201,DFT+U_failed_pakdel2025effect}. Indeed, this view is supported by recent first-principles calculations based on the constrained random phase approximation (cRPA) \cite{CRPA,CRPA2,aryasetiawan2006calculations,PhysRevMaterials.5.085001}, which reveal a pronounced, nonmonotonic, and chemically sensitive evolution of \(U\) across transition-metal perovskites \cite{cRPA_pero_PhysRevMaterials.9.015001}. In particular, systematic cRPA studies of 3$d$ ABO$_3$ compounds demonstrate a highly nonlinear dependence of \(U\) on $d$-electron filling, orbital hybridization, and screening channels, even within the same structural family~\cite{cRPA_pero_PhysRevMaterials.9.015001}. Such results have firmly established the belief that a uniform interaction strength is incompatible with DFT+$U$ and related static mean-field approaches. This non-transferability of \(U\) has motivated extensive efforts to compute interaction parameters explicitly for large materials sets, including recent high-throughput linear-response calculations of \(U\) on about thousand transition-metal-containing compounds for DFT+$U$ databases \cite{LR_HT_Ucal_PhysRevMaterials.8.014409}.

In contrast, the situation changes qualitatively within the embedded DMFT (eDMFT) framework~\cite{eDMFT_JPSJ}. Recent developments in constrained DMFT (c‑DMFT) allow the Coulomb interaction parameter to be determined self‑consistently within the same many‑body formalism used to compute the electronic structure~\cite{AS-cDMFT-PRBL-2026}. This approach naturally incorporates vertex corrections arising from local dynamical correlations, thereby treating electronic correlations and screening on an equal footing \cite{AS-cDMFT-PRBL-2026}. Applications of c-DMFT to a broad range of 3$d$ systems reveal a clear separation of interaction scales: metallic compounds typically yield \(U \approx 5\!-\!7\)~eV, while insulating systems require larger values of \(U \approx 8\!-\!10\)~eV. Importantly, the resulting spectral functions exhibit only weak sensitivity to moderate variations of \(U\) around these characteristic values—a behavior fundamentally distinct from DFT+$U$, where small changes in \(U\) can qualitatively alter electronic and magnetic solutions \cite{PhysRevB.58.1201,AS_NdVO4_DFTU,DFT+U_failed_pakdel2025effect,emery2017high, PhysRevB.60.15674, PhysRevB.73.195107}.

These observations motivate a hypothesis that fundamentally challenges the DFT+$U$ paradigm: namely, that fixed interaction strengths can reliably describe broad classes of correlated materials within the eDMFT framework. Specifically, here we hypothesize that uniform values of \(U=6\)~eV for $3d$-metals and \(U=10\)~eV for $3d$-insulators can capture the essential electronic structure of ABO$_3$ perovskite oxides when treated within eDMFT. This hypothesis must be validated directly against experimental photoemission measurements. If successful, it would eliminate the need for material-by-material parameter tuning and enable genuinely high-throughput, \textit{parameter-tuning-free} eDMFT calculations, opening a realistic pathway toward the construction of predictive electronic-structure databases tailored to strongly correlated materials.

In this work, we test this hypothesis through systematic high-throughput eDMFT calculations for ABO$_3$ perovskite oxides with A = Ca, Sr, La and B = V--Ni, using crystal structures obtained from the Crystallography Open Database (COD) \cite{COD1_Vaitkus2023,COD2_Grazulis2012,COD3_Grazulis2009}. By fixing the $A$ site and varying the $B$ site, and vice versa, we compute total densities of states and momentum-resolved spectral functions with fixed interaction strengths of \(U=6\)~eV for metallic compounds and \(U=10\)~eV for insulating compounds. The resulting spectra show overall excellent agreement with available photoemission data in the paramagnetic phase, with the exception of SrMnO$_3$ and LaCrO$_3$, where antiferromagnetic order is essential for quantitative agreement. Our results demonstrate that a fixed-$U$ eDMFT strategy is both accurate and practical, providing a viable foundation for future database development tailored to strongly correlated materials.

The rest of the paper is organized as follows. The computational method and structural details are described briefly in Sec. \ref{method} followed by results
and discussions in Sec. \ref{result}. Our main findings are summarized in Sec. \ref{conclusions}.

\begin{figure}[t]
\centering
\includegraphics[width=0.49\textwidth]{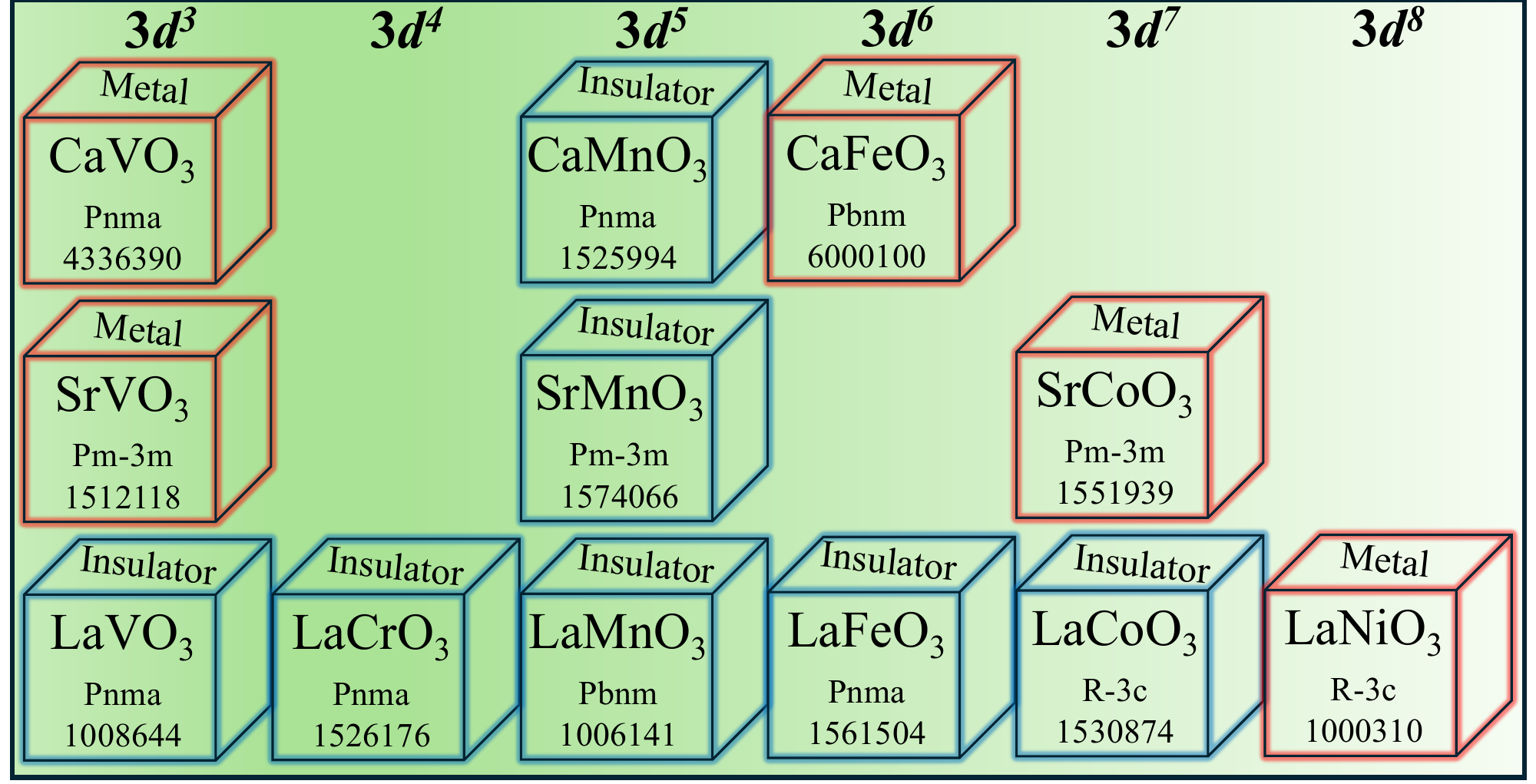}
\caption{Overview of the ABO$_3$ perovskite oxides investigated in this work. Shown are the experimentally reported crystal structures (space groups), Crystallography Open Database (COD) identification numbers, and metallic or insulating ground states with A = Ca, Sr, La and B = V, Cr, Mn, Fe, Co, Ni. The materials are organized according to the nominal transition-metal $3d$ electron count (left$\rightarrow$ right) fixing the A-site in a periodic table fashion, which also highlights the evolution from correlated metals to insulators across the series. }
\label{Fig1_materials}
\end{figure}

\begin{figure*}[t]
\centering
\includegraphics[width=1.0\textwidth]{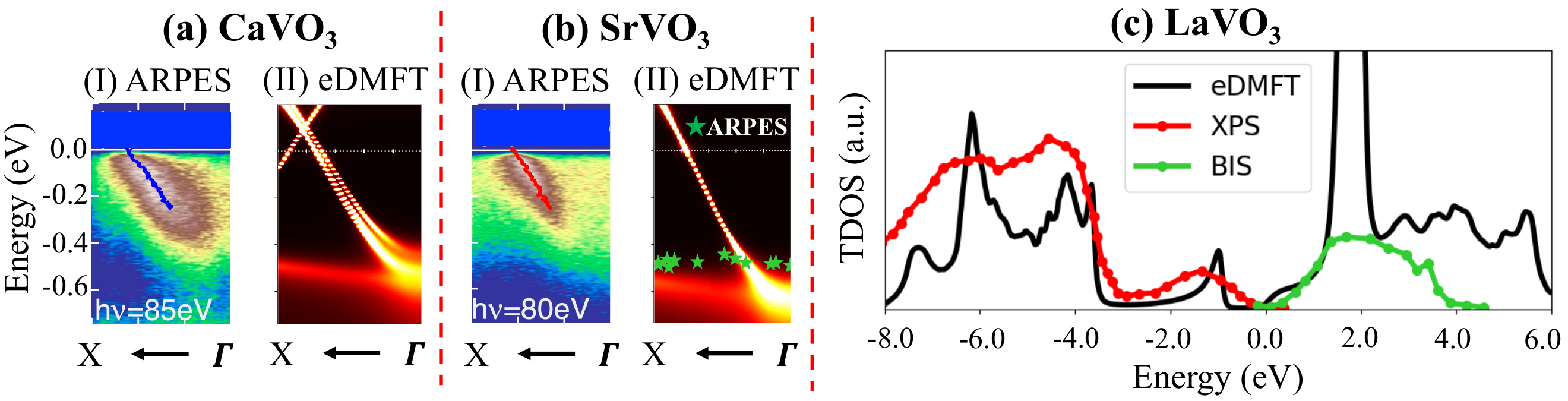}
\caption{Comparison of experimental and eDMFT spectral properties for vanadium-based perovskites. Panels (a) and (b) show momentum-resolved spectral functions of CaVO$_3$ and SrVO$_3$ along the $\Gamma$–$X$ direction. ARPES data [panels (I)] are reproduced from the literature \cite{yoshida2010mass}, with corresponding eDMFT results shown in panels (II). For SrVO$_3$, the second derivative of the ARPES energy distribution curves is overlaid as green stars \cite{takizawa2009coherent,yoshida2016correlated}. Panel (c) compares the eDMFT total density of states (TDOS, black) for LaVO$_3$ with experimental x-ray photoemission spectroscopy (XPS, red) and bremsstrahlung isochromat spectroscopy (BIS, green); experimental spectra are reproduced from Ref.~\cite{maiti2000spectroscopic}.}
\label{Fig2_avo3}
\end{figure*}

\section{Methods and Structural Details} \label{method}

We investigate twelve ABO$_3$ perovskite oxides, where A = Ca, Sr, La and B = V, Cr, Mn, Fe, Co, Ni, using experimental crystal structures obtained from the COD \cite{COD1_Vaitkus2023,COD2_Grazulis2012,COD3_Grazulis2009}. The corresponding COD identifiers, space groups, and metallic or insulating character of each compound are summarized in Fig.~\ref{Fig1_materials}. Experimentally reported antiferromagnetic (AFM) structures are taken from the Bilbao MAGNDATA database \cite{bilbaoMag_gallego2016magndata}. All paramagnetic (PM) and AFM calculations are performed using the density functional theory plus embedded dynamical mean-field theory (DFT+eDMFT) framework \cite{eDMFT_JPSJ,eDMFT2010} to compute spectral properties. The DFT component employs the Perdew--Burke--Ernzerhof (PBE) exchange--correlation functional and is implemented using the all-electron full-potential linearized augmented plane-wave method as implemented in the \textsc{WIEN2k} code \cite{PBE_PhysRevLett.77.3865,WIEN2k}. In the DFT+eDMFT formulation, the quantum impurity problem is solved using the continuous-time quantum Monte Carlo (CTQMC) method \cite{ctqmc}. All eDMFT calculations are carried out self-consistently on the Matsubara frequency axis, followed by analytical continuation to the real-frequency axis using the maximum entropy method to obtain spectral functions. High-throughput eDMFT simulations are performed at a temperature of $T = 230$~K ($\beta = 50$~eV$^{-1}$), using a Monkhorst--Pack $k$-point mesh of at least $10 \times 10 \times 10$ and a minimum of $4.5 \times 10^{9}$ Monte Carlo steps per iteration. The {\it exact} double-counting scheme~\cite{ExactDC} and density--density (Ising-type) Coulomb interaction are employed, with a fixed Hund’s coupling of $J = 0.8$~eV \cite{ExactDC}. A large energy window of $\pm 10$~eV around the Fermi level is used in all eDMFT calculations to accurately capture hybridization between localized and itinerant electronic states.

All ABO$_3$ perovskites considered in this work crystallize experimentally in the \textit{Pnma}, \textit{Pbnm}, \textit{Pm$\bar{3}$m}, or \textit{R$\bar{3}$c} space groups. As representative examples, self-consistently determined Coulomb interaction parameter \(U\) is obtained within the constrained-DMFT (cDMFT) framework following the procedure described in Ref.~\cite{AS-cDMFT-PRBL-2026}. Employing \(2 \times 2 \times 2\) supercells, for SrVO$_3$, SrMnO$_3$, and LaCrO$_3$, the resulting values of \(U\) are approximately 6.1~eV, 7.3~eV, and 9.8~eV, respectively. These compounds serve as illustrative cases from the broader dataset. The computed interaction strengths fall within the range established by earlier cDMFT studies, with typical values of 5–7~eV for correlated metals and 7–10~eV for correlated insulators~\cite{AS-cDMFT-PRBL-2026}. Hence, we consider $U$ = 6 eV  for all metals and $U$ = 10 eV for all insulators to compute the electronic structures of ABO$_3$ (where A = Ca, Sr, La and B = V, Cr, Mn, Fe, Co, Ni). Since no \textit{a priori} information is available on whether a given compound among the twelve ABO$_3$ is metallic or insulating for a specific crystal structure, the classification is based on reported transport and optical measurements available in the literature. We also note that matrix-element effects present in experimental photoemission spectra are not included in the eDMFT spectral functions; consequently, while peak positions and the overall distribution of spectral weight are meaningfully compared, quantitative agreement in peak intensities is not expected.

\begin{figure*}[t!]
\centering
\includegraphics[width=0.8\textwidth]{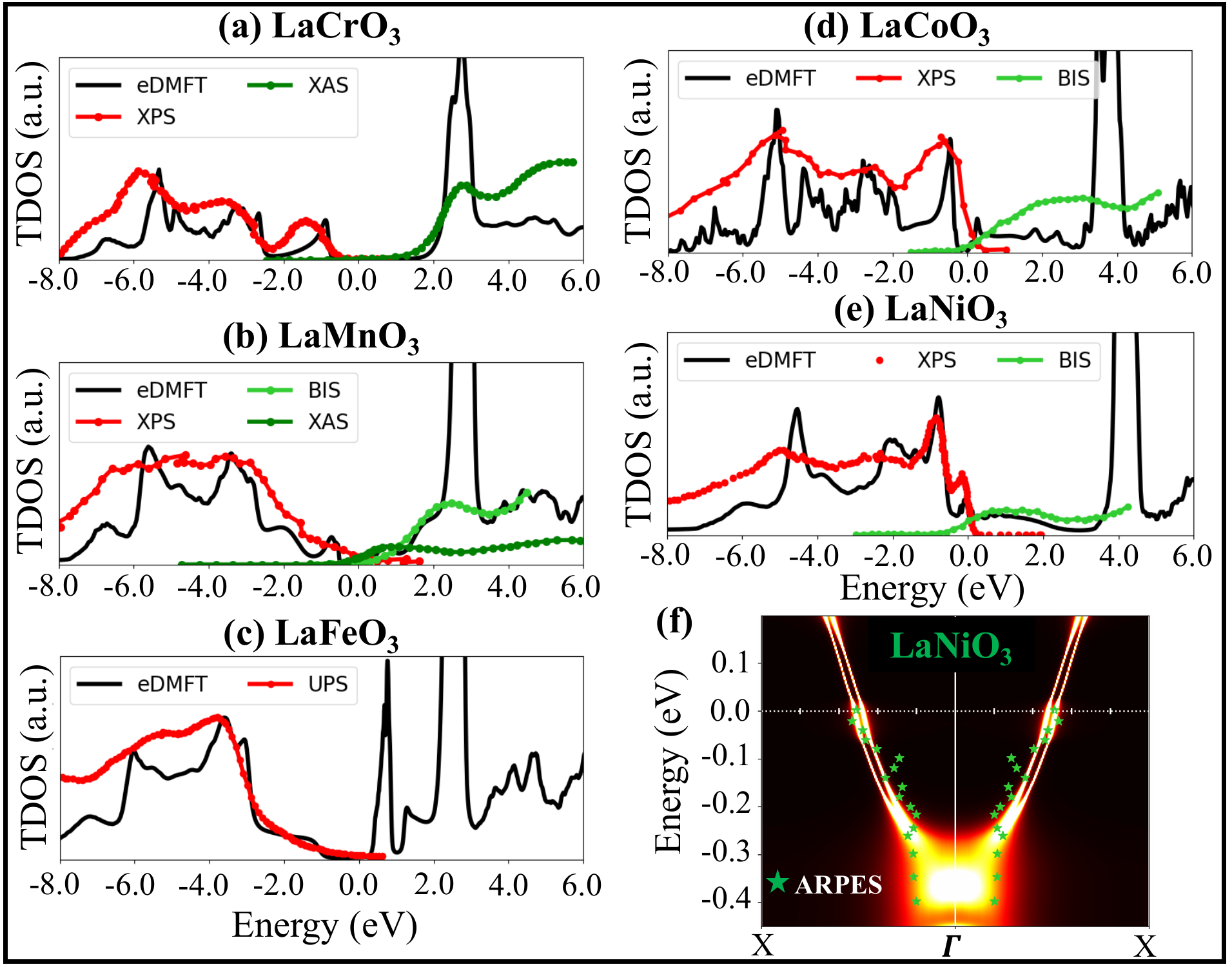}
\caption{Spectral properties of La$M$O$_3$ ($M$ = Cr, Mn, Fe, Co, Ni) in the paramagnetic phase. Panels (a)--(e) show the eDMFT total density of states (TDOS, black) compared with experimental x-ray photoemission spectroscopy (XPS, red), x-ray absorption spectroscopy (XAS, green), and bremsstrahlung isochromat spectroscopy (BIS, lime green) data reproduced from the literature \cite{LaCrO3_PhysRevB.91.155129,LaMnO3_xps,LaMnO3_bis,LaMnO3_xas,LaFeO3_UPS,LaCoO3_XPS,LaNiO3_XPS}. Panel (f) compares the eDMFT momentum-resolved spectral function of metallic LaNiO$_3$ with angle-resolved photoemission spectroscopy (ARPES) data (green stars) reproduced from Ref.~\cite{LaNiO3_ARPES}.}
\label{Fig2_LaBo3}
\end{figure*}

\section{Results and Discussions} \label{result}

The results are organized as follows. In Sec.~\ref{avo}, we compare theoretical and experimental spectra for vanadium-based perovskites by varying the $A$ site (Ca, Sr, and La). The effect of $B$-site substitution is then examined in Sec.~\ref{laBo} through a systematic analysis of La-based oxides with $B$ = Cr, Mn, Fe, Co, and Ni. In Sec.~\ref{ca_srxo}, we discuss CaXO$_3$ (X = Mn, Fe) and SrXO$_3$ (X = Mn, Co) to further elucidate trends associated with alkaline-earth substitution. Finally, the role of magnetic ordering is addressed for SrMnO$_3$, LaCrO$_3$, and LaMnO$_3$ by comparing PM and AFM spectral properties using the same value of $U$.

\subsection{$A$VO$_3$ ($A$ = Ca, Sr, La)} \label{avo}

We first examine early 3{\it d} vanadium-based perovskite oxides with the $A$ site being Ca, Sr, and La in ABO$_3$. Here both CaVO$_3$ and SrVO$_3$ are well-known correlated metals, and LaVO$_3$ is known as a Mott–Hubbard insulator. Figures~\ref{Fig2_avo3}(a) and \ref{Fig2_avo3}(b) compare angle-resolved photoemission spectroscopy (ARPES) data with the corresponding eDMFT momentum-resolved spectral functions along the $\Gamma$--$X$ direction. SrVO$_3$ crystallizes in the ideal cubic perovskite structure, whereas CaVO$_3$ exhibits orthorhombic distortions that enhance electronic correlations. Experimentally, this manifests as a narrower bandwidth in CaVO$_3$ compared to SrVO$_3$, as seen by comparing Figs.~\ref{Fig2_avo3}(a)(I) and \ref{Fig2_avo3}(b)(I). A consistent trend is captured by the eDMFT spectra [Figs.~\ref{Fig2_avo3}(a)(II) and \ref{Fig2_avo3}(b)(II)], where a bandwidth reduction of approximately 0.1~eV at the $\Gamma$ point is obtained for CaVO$_3$ relative to SrVO$_3$. The ARPES-measured occupied bandwidth of SrVO$_3$ thin films at the $\Gamma$ point is $-0.44 \pm 0.02$~eV, in good agreement with the eDMFT value of approximately $-0.5$~eV \cite{takizawa2009coherent,yoshida2016correlated}. In addition, a dispersionless feature is observed in the eDMFT spectra along the $\Gamma$--$X$ direction at energies around $-0.5$~eV for SrVO$_3$ and $-0.45$~eV for CaVO$_3$. This is consistent with ARPES measurements on high-quality SrVO$_3$ thin films \cite{takizawa2009coherent,yoshida2016correlated}, which are marked with green star in Fig.\ref{Fig2_avo3}(b)-II.

We next turn to another vanadium oxide, LaVO$_3$, which is a prototypical Mott–Hubbard insulator. Figure~\ref{Fig2_avo3}(c) shows the total density of states (TDOS) obtained from eDMFT together with experimental spectra, where x-ray photoemission spectroscopy (XPS) probes the occupied states and bremsstrahlung isochromat spectroscopy (BIS) probes the unoccupied states. Since the experimental photoemission (PES) and inverse PIS (IPES) spectra are reported in arbitrary units, we have re-scaled them in y-direction to fit in the range of the computed TDOS. A Mott insulating gap of approximately 1~eV is obtained from the TDOS, in good agreement with the experimentally reported value \cite{maiti2000spectroscopic}. The lower Hubbard band appears at about $-1.2$~eV in the TDOS, consistent with the position of the main XPS peak. At higher binding energies beyond $\sim -3.5$~eV, broader features predominantly of O-$2p$ character are observed in both the TDOS and XPS spectra. For the unoccupied states, the TDOS reproduces well the peak positions observed in BIS. Overall, the major experimental spectral features of LaVO$_3$ are well captured by the eDMFT results, in agreement with previous theoretical studies \cite{KH_lvo_PhysRevB.90.075136, LaVO3_PhysRevLett.99.126402}.

\subsection{La$M$O$_3$ ($M$ = Cr, Mn, Fe, Co, Ni)} \label{laBo}

We next analyze the spectral properties of La$M$O$_3$ perovskites with systematic variation of the $B$ site ($M$ = Cr--Ni), as shown in Figs.~\ref{Fig2_LaBo3}(a)--(e). The eDMFT TDOS is computed in the PM phase and compared with experimentally measured spectra, where XPS probes the occupied states, and x-ray absorption spectroscopy (XAS) or BIS probes the unoccupied states.
\begin{figure}[t!]
\centering
\includegraphics[width=0.48\textwidth]{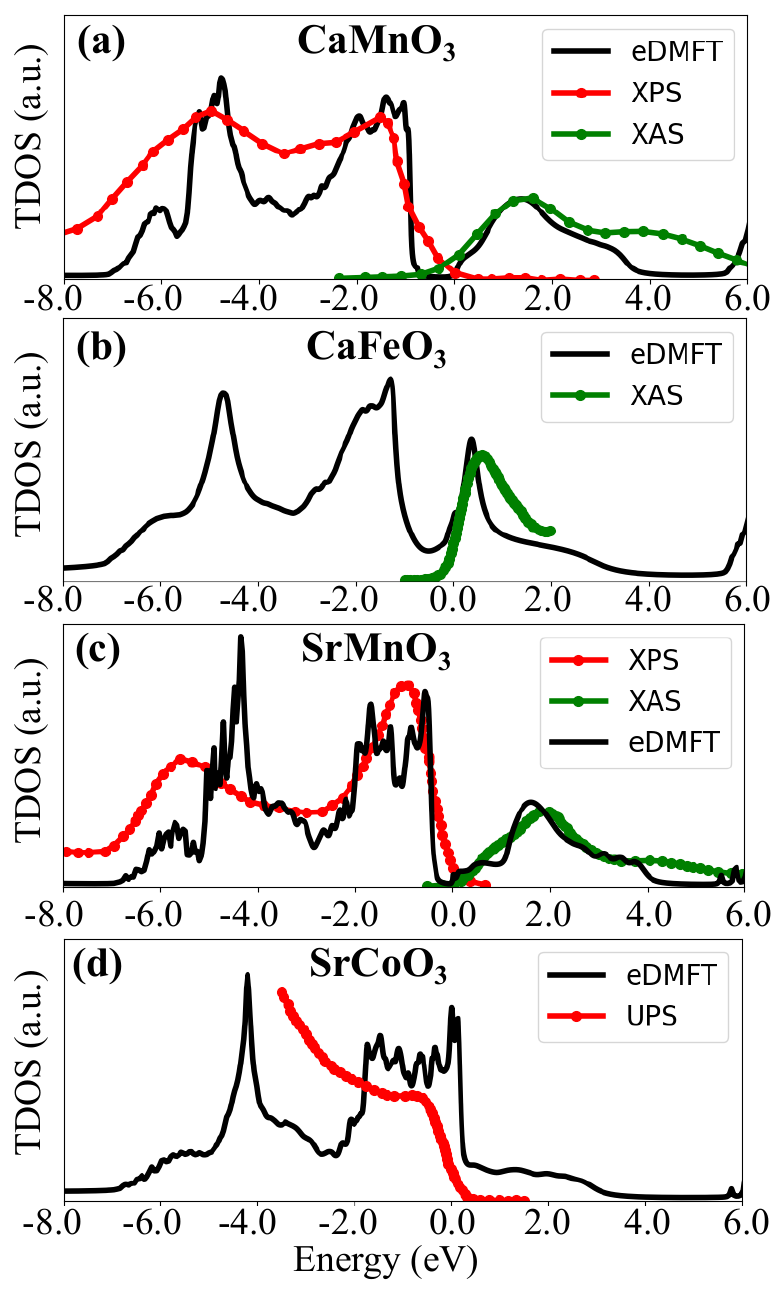}
\caption{Spectral properties of CaXO$_3$ (X = Mn, Fe) and SrXO$_3$ (X = Mn, Co) in the paramagnetic phase. Panels (a)--(d) show the eDMFT-computed total density of states (TDOS, black) compared with experimental x-ray photoemission spectroscopy (XPS, red) and x-ray absorption spectroscopy (XAS, green) data reproduced from the literature \cite{CaMnO3_XPS_PhysRevLett.76.4215,CaMnO3_zampieri2002xps,CaFeO3_PhysRevMaterials.2.015002,SrMnO3_PhysRevB.81.073101}. For SrCoO$_3$, the red curve denotes ultraviolet photoelectron spectroscopy (UPS) data reproduced from Ref.~\cite{SrCoO3_chowdhury2021spectroscopic}.}
\label{Fig2_sr_cabo3}
\end{figure}
Experimentally, photoemission spectroscopy (PES) provides higher energy resolution than inverse photoemission spectroscopy (IPES), while the latter typically exhibits larger spectral broadening due to instrumental limitations. Consequently, extracting and comparing a single numerical value for the band gap is not an appropriate metric for assessing agreement with the calculated TDOS. Instead, we perform a direct, feature-by-feature comparison between the theoretical and experimental spectra. In Fig. ~\ref{Fig2_LaBo3}, we compare eDMFT-computed TDOS in the PM phase for (a) LaCrO$_3$, (b) LaMnO$_3$, (c) LaFeO$_3$, (d) LaCoO$_3$ and (e) LaNiO$_3$, which are shown in black solid lines, while red, green and lime green dotted solid lines indicate XPS, XAS and BIS spectra as reproduced from Refs.~\cite{LaCrO3_PhysRevB.91.155129},~\cite{LaMnO3_xps,LaMnO3_bis},~\cite{LaMnO3_xas},~\cite{LaFeO3_UPS},~\cite{LaCoO3_XPS}, and ~\cite{LaNiO3_XPS} respectively. Notably, for LaCrO$_3$, the unoccupied eDMFT spectra exhibit good agreement with XAS without requiring a rigid energy shift of $\sim$1~eV, which is commonly applied when comparing first principles results with experiment \cite{LaCrO3_PhysRevB.91.155129}. For La$M$O$_3$ compounds with $M$ = Mn, Fe, Co, and Ni, the dominant peak positions in the eDMFT TDOS show good overall agreement with the corresponding experimental XPS, XAS, and BIS features. For LaCoO$_3$, both the experimental XPS and BIS spectra are broad. In contrast, LaCrO$_3$ exhibits a well-defined Mott gap both in experiment and in eDMFT. However, the occupied spectral features of LaCrO$_3$ are shifted toward higher energies by approximately 0.5~eV relative to the XPS data. This discrepancy is attributed to the use of PM eDMFT calculations at 230~K, whereas LaCrO$_3$ is an antiferromagnetic insulator below 380~K. The role of magnetic ordering in this compound is therefore examined separately in Sec.~\ref{magnetic}.

Nevertheless, the TDOS computed using a fixed interaction strength of $U=10$~eV reproduce the experimentally observed PES/IPES features within experimental uncertainty. Finally, in Fig.~\ref{Fig2_LaBo3}(f), we compare the eDMFT momentum-resolved spectral function along the $\Gamma$--$X$ direction with available ARPES data for metallic LaNiO$_3$~\cite{LaNiO3_ARPES_PhysRevB.79.115122}, where $U$ is reduced to 6 eV. The calculated dispersions and spectral weight distribution show a very good agreement with the experimental ARPES results, further validating the fixed-$U$ eDMFT description across the La$M$O$_3$ series.

\subsection{CaXO$_3$ (X = Mn, Fe) \& SrXO$_3$ (X = Mn, Co)} \label{ca_srxo}

We extend discussion to CaXO$_3$ (X = Mn, Fe) and SrXO$_3$ (X = Mn, Co), with the comparison between PM eDMFT spectral functions and available experimental spectra, XPS, XAS, and ultraviolet photoemission spectroscopy (UPS), as shown in Figs.~\ref{Fig2_sr_cabo3}(a)--(d). Overall, the energy positions of the dominant peaks in the eDMFT TDOS are in good agreement with the corresponding experimental features for all compounds. For CaMnO$_3$, the XPS spectra exhibit two prominent peaks at approximately $-1.0$~eV (P1) and $-5.5$~eV (P2), both of which are well reproduced by the eDMFT TDOS. A similar peak structure is observed for SrMnO$_3$, with experimental P1 and P2 located at approximately $-0.5$~eV and $-4.4$~eV, respectively. In this case, the agreement between PES and eDMFT is less quantitative, as the corresponding eDMFT peaks appear at slightly shifted energies. This discrepancy is attributed to magnetic odering effects, since SrMnO$_3$ undergoes antiferromagnetic ordering below its N\'eel temperature of 233--260~K \cite{takeda1974magnetic,kikuchi1999syntheses,PhysRevB.64.134412, PhysRevB.74.144102}, whereas the present calculations are performed in the PM phase. For metallic CaFeO$_3$, high-quality PES data are not available to the best of our knowledge; therefore, the comparison is restricted to inverse photoemission spectra. As shown in Fig.~\ref{Fig2_sr_cabo3}(b), the eDMFT TDOS reproduces the experimental IPES features well.  It is worth noting that our analysis is restricted to the orthorhombic $Pbnm$ phase of CaFeO$_3$, which is metallic; the high-temperature insulating $P2_1/n$ phase is not considered in the present study. For Hund's metallic SrCoO$_3$, the IPES data are, to the best of our knowledge, unavailable. For the PES spectra, we are able to extract experimental  ultraviolet photoelectron spectroscopy (UPS) data (Ref.~\cite{SrCoO3_chowdhury2021spectroscopic}) only up to the onset of the second peak, which exhibits a broad spectral feature and compare reasonably with eDMFT.

\subsection{Effect of magnetic ordering} \label{magnetic}

\begin{figure}[t]
\centering
\includegraphics[width=0.48\textwidth]{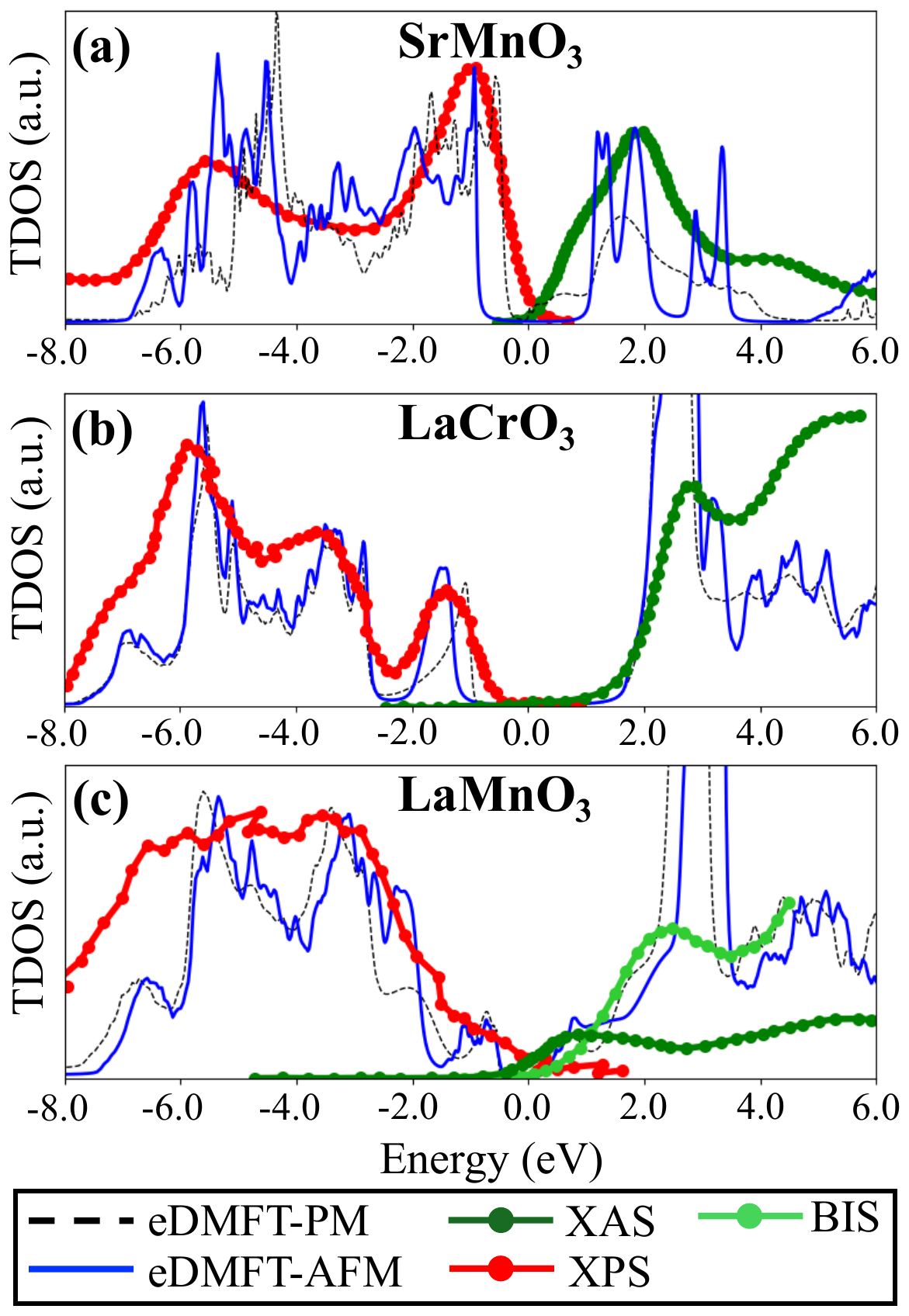}
\caption{Effect of magnetic ordering on the spectral properties of selected ABO$_3$ perovskites. Panels (a)--(c) show the eDMFT total density of states (TDOS) for SrMnO$_3$, LaCrO$_3$, and LaMnO$_3$, respectively, computed in the paramagnetic (PM, black dashed lines) and antiferromagnetic (AFM, blue solid lines) phases. Experimental x-ray photoemission spectroscopy (XPS, red), x-ray absorption spectroscopy (XAS, dark green), and bremsstrahlung isochromat spectroscopy (BIS, light green) data are reproduced from the literature \cite{SrMnO3_PhysRevB.81.073101,LaCrO3_PhysRevB.91.155129,LaMnO3_xps,LaMnO3_xas,LaMnO3_bis}.}
\label{fig_mag}
\end{figure}

We now examine the role of magnetic ordering in ABO$_3$ perovskites by explicitly considering AFM phases and comparing the resulting spectra with those obtained in the PM phase. Experimentally determined magnetic structures for SrMnO$_3$, LaCrO$_3$, and LaMnO$_3$ are taken from the open-access Bilbao MAGNDATA database \cite{bilbaoMag_gallego2016magndata}. SrMnO$_3$ and LaCrO$_3$ are considered in the G-type AFM configuration, while LaMnO$_3$ is treated in the experimentally observed A-type AFM phase. Among the materials investigated in this work, only these three compounds have experimentally resolved magnetic structures available in the database. To isolate the effect of magnetic ordering, the same interaction strength of $U=10$~eV is employed in both PM and AFM eDMFT calculations.
The eDMFT-computed TDOS for SrMnO$_3$, LaCrO$_3$, and LaMnO$_3$ in both PM and AFM phases are shown in Figs.~\ref{fig_mag}(a)--(c), together with the corresponding experimental XPS, XAS, and BIS spectra. In charge-transfer insulators, such as binary transition-metal oxides, the insulating gap is often relatively insensitive to magnetic ordering because the low-energy excitations involve substantial oxygen--transition-metal hybridization rather than purely local electronic states \cite{TMO2-SM}. In contrast, magnetic ordering plays a much more pronounced role in the present Hund-stabilized correlated insulators, particularly for SrMnO$_3$ and LaCrO$_3$.
For SrMnO$_3$, the AFM eDMFT TDOS shows markedly improved agreement with the experimental XPS and XAS peak positions relative to the PM results. The insulating gap extracted from the AFM TDOS is approximately 1.4~eV, which is much larger than the gap in the PM phase. A similar improvement is observed for LaCrO$_3$, as shown in Fig.~\ref{fig_mag}(b), where AFM ordering yields a substantially better reproduction of the occupied spectral features compared to PM eDMFT. The resulting gap for AFM-LaCrO$_3$ is approximately 2.5~eV, consistent with the experimentally reported range of 2.7--2.8~eV in the XAS-BIS experiments~\cite{LaCrO3_gap_PhysRevB.54.7816,LaCrO3_gap1_PhysRevLett.110.077401}.

In contrast, for LaMnO$_3$ [Fig.~\ref{fig_mag}(c)], both PM and AFM eDMFT TDOS show similarly good agreement with the experimental spectra, indicating that its spectral properties are comparatively less sensitive to magnetic ordering. Overall, these results demonstrate that incorporating experimentally observed magnetic order, while retaining the same interaction strength of $U=10$~eV, substantially improves the agreement between eDMFT spectral functions and experiment for SrMnO$_3$ and LaCrO$_3$. This behavior highlights the important role of magnetic correlations in shaping the low-energy electronic structure of these correlated insulating oxides.

\section{Conclusions} \label{conclusions}

In this work, we carried out a systematic high-throughput investigation of the electronic structure and spectral properties of ABO$_3$ perovskite oxides using charge self-consistent eDMFT. By examining a broad set of materials spanning correlated metals and correlated insulators, we demonstrate that physically motivated interaction values of \(U=6\)~eV for metallic compounds and \(U=10\)~eV for insulating compounds provide a quantitatively reliable description of spectral properties within the eDMFT framework. These interaction scales are consistent with recent constrained-eDMFT calculations for broad classes of $3d$ transition-metal oxides and stand in sharp contrast to the conventional DFT+$U$ paradigm, where the interaction strength is generally assumed to be strongly material dependent.

The eDMFT calculations reproduce key experimental observables across diverse chemical compositions and crystal structures, including bandwidth renormalization, Hubbard-band formation, insulating gaps, and momentum-resolved dispersions. In particular, comparisons with photoemission, inverse photoemission, x-ray absorption, ultraviolet photoemission, and angle-resolved photoemission measurements show overall good agreement in spectral peak positions and low-energy electronic structure for vanadium-, chromium-, manganese-, iron-, cobalt-, and nickel-based perovskites. For compounds such as SrMnO$_3$ and LaCrO$_3$, incorporating experimentally observed antiferromagnetic order substantially improves agreement with experiment while retaining the same interaction strength as in the paramagnetic phase.

The reduced sensitivity of eDMFT spectral properties to moderate variations of \(U\) originates from the self-consistent treatment of electronic correlations and screening within the large-energy-window eDMFT framework employed here. In conventional downfolded DFT+DMFT approaches based on material-dependent Wannier orbitals, the effective low-energy Hamiltonian and corresponding interaction parameters can vary substantially across compounds, often requiring repeated material-specific determination or tuning of \(U\). In contrast, the present eDMFT implementation employs highly localized orbitals spanning a large energy window, allowing a substantial portion of the screening processes to be treated internally and self-consistently within the many-body framework itself. As a result, the low-energy spectral properties are governed primarily by the dynamical self-energy rather than by static interaction-induced energy shifts, leading to a remarkable transferability of the interaction scales across chemically distinct compounds. Taken together, our results establish that {\it parameter-tuning-free} eDMFT calculations can provide predictive electronic structures for a broad class of correlated perovskite oxides, where the Coulomb interaction $U$ can be systematically assigned for $3d$ metals and insulators. More broadly, this work demonstrates the feasibility of scalable, high-throughput many-body calculations for correlated materials and opens a practical pathway toward next-generation correlated-electron materials databases beyond conventional DFT-based frameworks.

\section{Acknowledgments}

Authors acknowledge the support from the National Science Foundation Award No. NSF OAC-2311557 and NSF OAC-2311558. Authors benefited from the Frontera supercomputer at the Texas Advanced Computing Center (TACC) at The University of Texas at Austin, which is supported by National Science Foundation Grant No. OAC-1818253.
KH also acknowledges support of NSF DMR-2233892. \\

\section{Data availability}
All data that support the plots within this paper and other findings of this study are available from the corresponding author upon reasonable request. Source data are provided with this manuscript.

%

\end{document}